\documentclass{article}

\usepackage{enumitem}
\usepackage{authblk}

\usepackage{graphicx} 
\usepackage{svg}
\usepackage[square,numbers]{natbib}
\usepackage{float}

\usepackage{comment}

\usepackage{url}

\providecommand{\keywords}[1]
{
  \small	
  \textbf{\textit{Keywords---}} #1
}
\begin{document}

\title{Measuring Geographic Diversity of Foundation Models with a Natural Language--based Geo-guessing Experiment on GPT-4}

\author[1]{Zilong Liu}
\author[1,2]{Krzysztof Janowicz}
\author[2]{Kitty Currier}
\author[1]{Meilin Shi}

\affil[1]{Department of Geography and Regional Research, University of Vienna, Austria}
\affil[2]{Department of Geography, University of California, Santa Barbara, USA}
\date{}

\maketitle
\begin{abstract}
Generative AI based on foundation models provides a first glimpse into the world represented by machines trained on vast amounts of multimodal data ingested by these models during training. If we consider the resulting models as knowledge bases in their own right, this may open up new avenues for understanding places through the lens of machines. In this work, we adopt this thinking and select GPT-4, a state-of-the-art representative in the family of multimodal large language models, to study its geographic diversity regarding how well geographic features are represented. Using DBpedia abstracts as a ground-truth corpus for probing, our natural language--based geo-guessing experiment shows that GPT-4 may currently encode insufficient knowledge about several geographic feature types on a global level. On a local level, we observe not only this insufficiency but also inter-regional disparities in GPT-4's geo-guessing performance on UNESCO World Heritage Sites that carry significance to both local and global populations, and the inter-regional disparities may become smaller as the geographic scale increases. Morever, whether assessing the geo-guessing performance on a global or local level, we find inter-model disparities in GPT-4's geo-guessing performance when comparing its unimodal and multimodal variants. We hope this work can initiate a discussion on geographic diversity as an ethical principle within the GIScience community in the face of global socio-technical challenges.
\end{abstract}

\section{Introduction}

Like humans, machines are capable of learning from observations to draw inferences. However, if we do not fully understand the components and nature of the geo-data landscape, naively feeding these data to machines for training, validation, and testing purposes could yield unexpected and undesired results. In a pioneering work in image classification, \cite{46553} conducted a stress test on the generalizability of two classifiers pre-trained on two of the most commonly used image benchmark datasets. For images crowdsourced from Hyderabad, India, neither classifier could recognize well categories like \textit{groom} and \textit{bridegroom}. Also, the classifier trained on one dataset showed poorer performance on web images from the Global South, e.g., Ethiopia. Such failures could be attributed to a more \textit{Western} representation bias exhibited by both benchmark datasets. Situating GIScience in the current AI4Science\footnote{\url{https://www.microsoft.com/en-us/research/lab/microsoft-research-ai4science}} trend, we must ask ourselves: Are these models being developed and used for knowledge discovery for the benefit of all, irrespective of where we are or where we come from \citep{janowicz2023philosophical}?

The issues of geographic diversity exist not only in computer vision, but also in natural language processing tasks such as geoparsing~\citep{liu2022geoparsing}. Just as we have realized this fact, the geo-data landscape is facing a disruption brought by the release of ChatGPT as a recent breakthrough in foundation models~\citep{bommasani2021opportunities}. More recent large language models (LLMs) also support modalities such as images, greatly improving text-to-image generation and visual question answering. This success in multimodality is significant for the next generation of GeoAI models that could also be pre-trained with geo-data ranging from location descriptions to remote sensing and street-level images, and from vector data to cartographic maps. However, such models would still suffer from a lack of geographic diversity when learning latent spatial representations in a task-agnostic manner~\citep{mai2022towards}. Additionally, more and more geo-data could become generated by machines at scale. On HuggingFace, there are 43,616, 14,864, and 354 models for text, text-to-image, and image-to-text generation, respectively\footnote{Retrieved from \url{https://huggingface.co/models} on January 10, 2024}, which can be further deployed and fine-tuned for various purposes. Currently, it costs only \$0.00025/1k characters for inputs and 20 times the price for outputs when using Gemini Pro, one of the state-of-the-art multimodal closed-source models\footnote{\url{https://blog.google/technology/ai/gemini-api-developers-cloud}}. The increasing accessibility of generative AI may foster a feedback loop, where content created by these models is used to train subsequent generations. This raises concerns about the potential to perpetuate and amplify biases present in current and future models.

In this short paper, we examine the geographic diversity---or lack thereof---of GPT-4\footnote{\url{https://openai.com/research/gpt-4}}, the state-of-the-art multimodal LLM in OpenAI's GPT series. \cite{janowicz2023philosophical} suggested that what an LLM reveals is a mirror of the world through multiple distortions, e.g., one from our observed world to the digital world and another from the sampled world to the learned (and possibly debiased) world, embedded in high-dimensional vector space. Our work uses this analogy to guide the investigation into the geographic diversity of GPT-4, in the process examining what it means for a foundation model to be called geographically diverse. The main subject of our investigation is the collection of \textbf{geographic features}\footnote{\url{https://wiki.gis.com/wiki/index.php/Geographic_feature}} that constitute gazetteers referred to as the \textit{vocabulary} of geography~\citep{jackson2006thinking}. This subject is different from previous studies that may fall into an environmental-determinism trap, as they tend to attribute local machine-learning failures simply to data bias against a studied area. Also, previous work ignores the modifiable areal unit problem~\citep{openshaw1984modifiable}, most often using country-level differences in data distribution and model performance as the sole indicator of geographic diversity. Stemming from the \textit{platial} root of GIScience, we consider that the notion of geographic diversity has another facet, i.e., how well geographic features are represented. These features could be areas where a concept holds true but shifts, physical features that extend across the landscape, or human-made sites that carry historical and cultural meaning. In addition to countries, other kinds of relevant geographical units could be used when assessing geographic diversity.

We approach this notion of geographic diversity centered around the \textit{extension} (i.e., the instances to which a category applies) of geographic feature types, and we believe it is necessary not only to record where models would fail but also to develop innovative ways of assessing geographic diversity. Therefore, we design a natural language--based geo-guessing experiment, and suggest using its performance as an indicator. During the experiment, we mask the geographic feature mentioned in a piece of text and ask GPT-4 to supply its actual name.

\section{Related work}
\cite{zhao2021deep} were among the first to try to theorize about the intersection of generative AI, GIScience and the broader discipline of geography. They raised the problem of \textit{deep fake geography}, which situates fake geography (e.g., location spoofing or the fact that maps could tell lies) in the deep-learning era, and conducted an empirical study by using generative adversarial networks to inject landscape features from two other cities into satellite images of Tacoma in Washington, United States. As the resulting images appear to be authentic, the authors later developed detection models using visual and frequency-domain features. In the same work, it was also predicted that deep fakes would become an inevitable part of our society, and therefore, how to understand the fast emergence and negative impacts of associated techniques remains a key question.

Interestingly, the rapid progress in LLMs makes it important to look at generative AI as not merely a data generator but as a knowledge base. \cite{petroni2019language} conducted a fill-in-the-blank cloze test on a wide range of pre-trained language models including BERT\footnote{\url{http://ai.googleblog.com/2018/11/open-sourcing-bert-state-of-art-pre.html}}, an early language model using the Transformer architecture which forms the fundamental building block of today's LLMs. They found that BERT can store relational knowledge in its training data and recall factual, commonsense knowledge without fine-tuning.

More recent work that involves knowledge extraction indicates that geographic knowledge, as a kind of specialized knowledge, is encoded in these models, as well. \cite{lietard2021language} designed three probing tasks about coordinates, population sizes, and neighboring countries, respectively. As the model size increased, more geographic knowledge was found to be learned. Similarly, \cite{bhandari2023large} focused directly on LLMs and probed for coordinates of cities. They found that LLaMA\footnote{\url{https://ai.meta.com/blog/large-language-model-llama-meta-ai}} in zero-shot settings can outperform LLaMA in few-shot settings. In addition, they discovered that LLMs have the ability to predict a place based on contextual information (containing an input place and a spatial preposition) and to achieve distance-based spatial reasoning about cities. \cite{jang2023understanding} retrieved textual responses (structured as bullet points) from ChatGPT and street-level images from DALL·E 2\footnote{\url{https://openai.com/dall-e-2}} to study the place identity of 31 cities. Then, they examined the semantic similarity between the place identity from the perspective of the models and the place identity embedded in two ground-truth text and image datasets. The results showed that ChatGPT and DALL·E 2 can represent salient features of cities. 

These post-BERT works suggest that the usage of generative AI in the form of LLMs should not be limited to content generation. Using GPT-4 as an example, we focus on its learned representation (rather than reasoning) about geographic features beyond administrative features, e.g., cities or countries. We probe it for factual knowledge in the form of unstructured texts rather than triples. In addition, our experiment differs from mainstream probing techniques that query about feature attributes. Instead, we query GPT-4 about a feature, itself, based on the assumption that contextual words are geo-indicative.

\section{Ground-Truth Data Acquisition}
Our ground-truth corpus is retrieved via SPARQL queries from DBpedia\footnote{\url{https://www.dbpedia.org}}. DBpedia is currently one of the largest open knowledge bases that uses Semantic Web and Linked Data technologies to extract structured data from Wikipedia~\citep{lehmann2015dbpedia}. We select geographic features that belong to subclasses of the \texttt{dbo:Place} category and subsequent subclasses, as well. This selection includes a subset of all geographic features that exist in DBpedia, in which other classes, such as \texttt{dbo:ArchitecturalStructure}, also contain relevant features.

\begin{figure*}[h!]
    \centering
    \caption{The retrieval process of a \texttt{dbo:Sea} feature \texttt{dbr:Mediterranean\_Sea} and its abstract from DBpedia}
    \includegraphics[width=\textwidth]{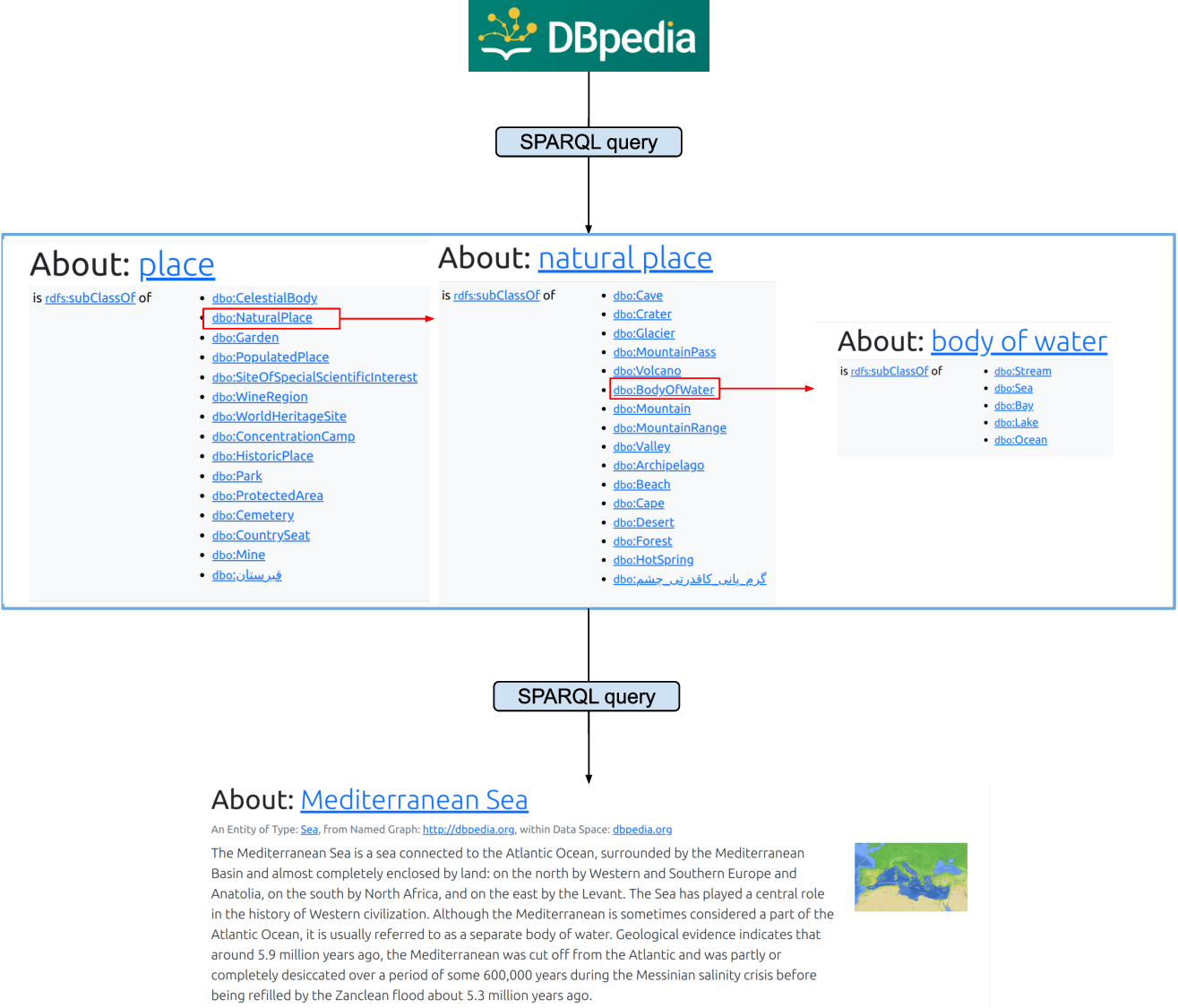}
    \label{fig:dbpedia_retrieval}
\end{figure*}

As our work does not explicitly involve the probing of multilingual knowledge of GPT-4, we retrieve only English abstracts which, however, may contain non-English feature names. Features that lack an English abstract and that lack mentions of their names in the abstract are omitted from our study. These additional classes are not considered in this work. Figure~\ref{fig:dbpedia_retrieval} shows the retrieval workflow, in which the first step is to retrieve \texttt{dbo:Place} subclasses and subsequent subclasses, and the second step is to retrieve the name and the abstract of an instance.

\section{Geo-guessing Experiment}
\begin{figure*}[h!]
    \centering
    \caption{An example geo-guessing experiment about a \texttt{dbo:Bay} feature \texttt{dbr:Gulf\_of\_Thailand}, implemented with the \texttt{Chat} mode in OpenAI Playground}
    \includegraphics[width=\textwidth]{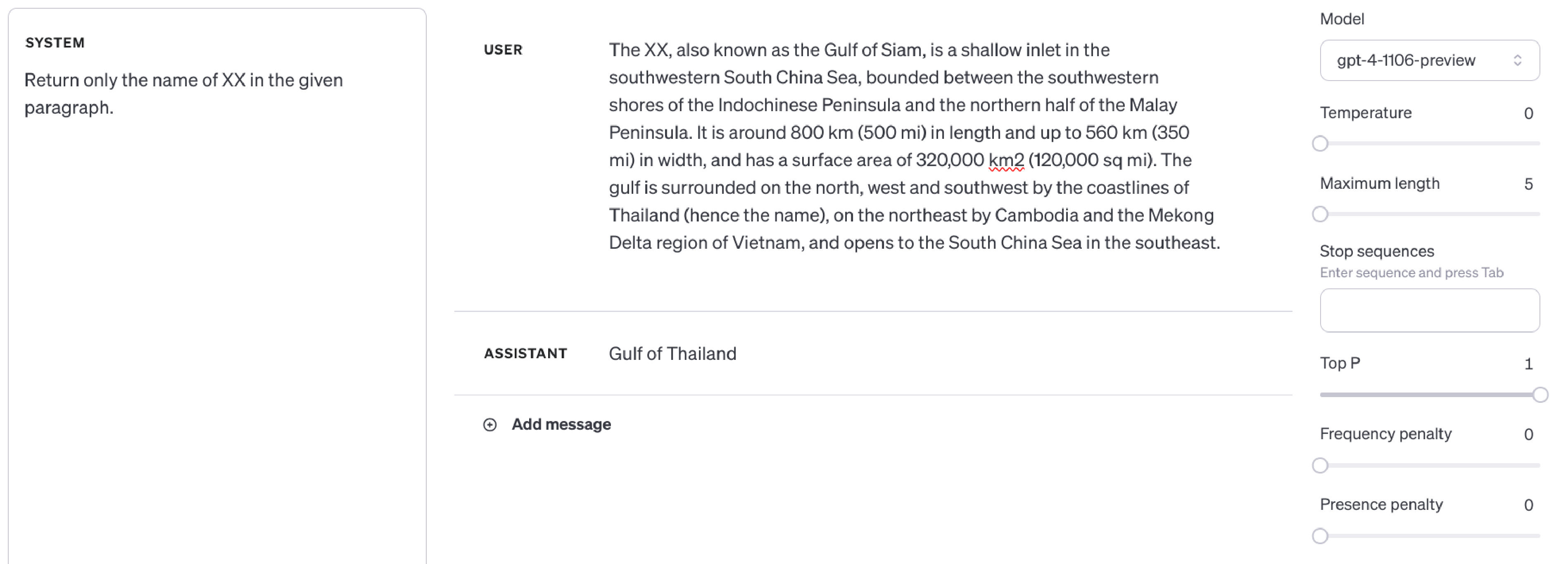}
    \label{fig:probing_example}
\end{figure*}

DBpedia abstracts allow us to conduct a geo-guessing experiment on GPT-4. We use two GPT-4 variants, \texttt{gpt-4-1106-preview} and \texttt{gpt-4-vision-preview}. Both models were trained with data up to April 2023. Compared with the \texttt{gpt-4-1106-preview} (that was the GPT-4 Turbo model before the more recent release of \texttt{gpt-4-0125-preview}), \texttt{gpt-4-vision-preview} has the additional ability to understand images, and therefore, \texttt{gpt-4-vision-preview} is multimodal. We probe both models in zero-shot settings and set the temperature (i.e., the randomness in the output) to 0. No candidate answer is provided for the model in the experiment, meaning that it is an open-ended question-answering task. Figure~\ref{fig:probing_example} shows an example of how the experiment can be achieved in the OpenAI Playground\footnote{\url{https://platform.openai.com/playground}}. The system prompt is \texttt{Return only the name of XX in the given paragraph}. The user prompt is an abstract that masks the target feature as \texttt{XX}. In this example, \texttt{gpt-4-1106-preview} outputs \texttt{Gulf of Thailand} as the correct answer. It is also worth noting that as GPT-4 uses \textit{both publicly available data (such as Internet data) and data licensed from third-party providers}~\citep{achiam2023gpt}, its training data may include DBpedia as an open knowledge source. Therefore, we assume that GPT-4 should output the precisely correct answer if it memorizes the corresponding parts of its training data.

\section{Evaluation Results}
\begin{figure*}[h!]
    \centering
    \caption{The hierarchy of DBpedia's \texttt{dbo:Place} subclasses used in our work-in-progress}
    \includegraphics[width=\textwidth]{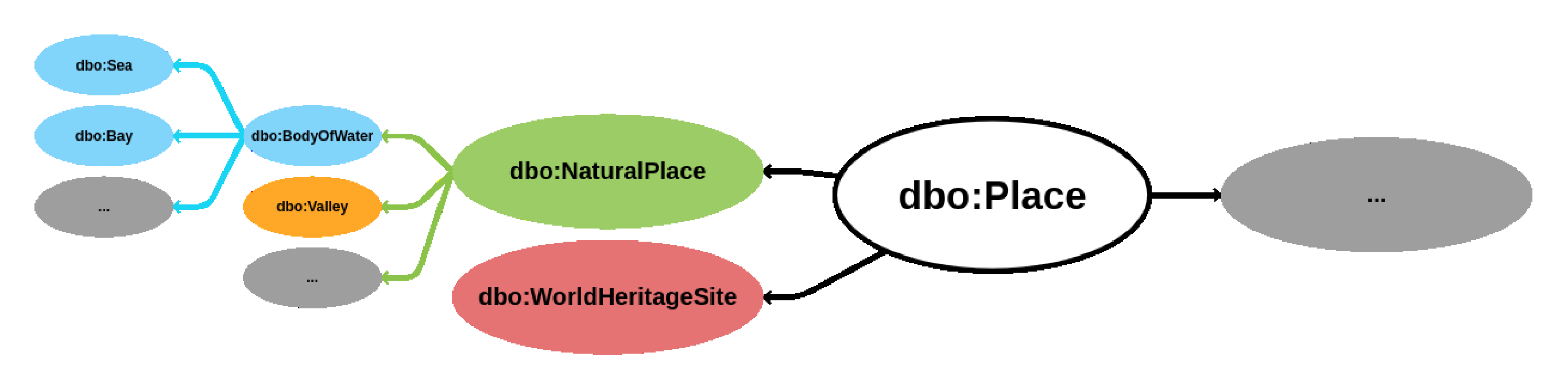}
    \label{fig:dbpedia_hierarchy}
\end{figure*}

Our current experiment involves four subclass types, \texttt{dbo:Valley}, \texttt{dbo:Bay}, \texttt{dbo:Sea}, and \texttt{dbo:WorldHeritageSite}. The feature types \texttt{dbo:Bay} and \texttt{dbo:Sea} are subclasses of \texttt{dbo:BodyOfWater}, and \texttt{dbo:BodyOfWater} and \texttt{dbo:Valley} are subclasses of \texttt{dbo:NaturalPlace}. Both \texttt{dbo:NaturalPlace} and \texttt{dbo:WorldHeritageSite} are subclasses of \texttt{dbo:Place}. Figure~\ref{fig:dbpedia_hierarchy} illustrates a DBpedia geographic feature-type hierarchy, in which the grey circle represents \texttt{dbo:Place} subclasses excluded from our current work. In total, there are 15 \texttt{dbo:Valley}, 40 \texttt{dbo:Bay}, 152 \texttt{dbo:Sea}, and 981 \texttt{dbo:WorldHeritageSite} instances, respectively, used in our experiment.

\subsection{Analysis Results on a Global Level}
First, we measured the geo-guessing performance as the percentage of features correctly named by GPT-4. Table~\ref{tab:eval_results} shows the evaluation results by model and feature type. For each feature type, \texttt{gpt-4-vision-preview} correctly predicted fewer than half of the total features. The model \texttt{gpt-4-1106-preview} correctly predicted slightly more than half of the features belonging to \texttt{dbo:Bay} and \texttt{dbo:Sea}, with values of 0.55 and 0.51, respectively. Both \texttt{gpt-4-1106-preview} and \texttt{gpt-4-vision-preview} performed lowest on \texttt{dbo:Valley} and highest on \texttt{dbo:Bay}. Except for \texttt{dbo:Valley} (0.2 versus 0.27), surprisingly, \texttt{gpt-4-1106-preview} outperformed \texttt{gpt-4-vision-preview} on three other feature types. This may indicate that a \texttt{gpt-4-vision-preview} trained on additional image data (e.g., image--text pairs) does not necessarily encode more geographic knowledge than the pure language model \texttt{gpt-4-1106-preview}.

\begin{table*}[h!]
    \caption{The percentage of correct predictions by \texttt{gpt-4-1106-preview} and \texttt{gpt-4-vision-preview} on features belonging to \texttt{dbo:WorldHeritageSite}, \texttt{dbo:Valley}, \texttt{dbo:Bay}, and \texttt{dbo:Sea}}
    \centering    \begin{tabular}{|c|c|c|} \hline 
         \textbf{Feature Types} &  \textbf{gpt-4-1106-preview}& \textbf{gpt-4-vision-preview}\\ \hline 
         dbo:WorldHeritageSite &  0.38& 0.31\\ \hline 
         dbo:Valley &  0.2& 0.27\\ \hline 
         dbo:Bay &  0.55& 0.475\\ \hline 
         dbo:Sea &  0.51& 0.46\\\hline
    \end{tabular}
    \label{tab:eval_results}
\end{table*}
\subsection{Local-Analysis Results about UNESCO World Heritage Sites}
From the four selected feature types, we focus on \texttt{dbo:WorldHeritageSite} features next. According to the United Nations Educational, Scientific and Cultural Organization (UNESCO), "World Heritage sites belong to all the peoples of the world, irrespective of the territory on which they are located"\footnote{\url{https://whc.unesco.org/en/about}}. Therefore, these sites are geographic features that carry both interpretations by local populations and universal values for all of humanity. Compared with the previous analysis on \texttt{dbo:WorldHeritageSite} features from a \textit{global} perspective, here we define localness with two kinds of geographical units to examine GPT-4's performance regarding this unique feature type. One kind of unit is countries, and the other one is regions defined by UNESCO for its activities\footnote{\url{https://whc.unesco.org/en/activities}}. We then measured GPT-4's performance as the percentage of correct predictions aggregated by these two units. When assessing by countries, we only include countries with more than ten sites in our ground-truth corpus.

Table~\ref{tab:eval_results_countries_whsites} shows the top ten countries ordered by the percentage of correct predictions by \texttt{gpt-4-1106-preview} and \texttt{gpt-4-vision-preview}, respectively, on \texttt{dbo:WorldHeritageSite} features. For both models, there were inter-country disparities in their geo-guessing performance. In addition, their performance was less than or equal to 0.5, indicating a severe lack of encoded knowledge in GPT-4 about \texttt{dbo:WorldHeritageSite} on a country level. Both models had the same accuracy for four countries, including Spain (0.31), Germany (0.23), Switzerland (0.17), and Chile (0.07). However, \texttt{gpt-4-1106-preview} performed better in all the rest six countries, including France (0.5 versus 0.2), India (0.47 versus 0.41), China (0.39 versus 0.33), Italy (0.38 versus 0.29), Belgium (0.33 versus 0.25), and Japan (0.19 versus 0.13). The inter-model disparities indicate that \texttt{gpt-4-1106-preview} generally had a better country-level performance than \texttt{gpt-4-vision-preview} when geo-guessing \texttt{dbo:WorldHeritageSite} features.

Table~\ref{tab:eval_results_regions_whsites} shows the UNESCO-regions ordered by the percentage of correct predictions by \texttt{gpt-4-1106-preview} and \texttt{gpt-4-vision-preview}, respectively, on \texttt{dbo:WorldHeritageSite} features. In addition to inter-UNESCO-regional disparities in the performance of both models, we again observe that their performance was less than 0.5, which indicates a similar lack of UNESCO-region-level knowledge about \texttt{dbo:WorldHeritageSite} encoded in GPT-4. Except for Arab States (0.28), \texttt{gpt-4-1106-preview} had a better performance than \texttt{gpt-4-vision-preview} in all the rest four UNESCO regions, including Latin America and the Caribbean (0.413 versus 0.26), Asia and the Pacific (0.407 versus 0.36), Africa (0.4 versus 0.37), and Europe and North America (0.36 versus 0.27). Again, this reveals inter-model disparities in GPT-4's geo-guessing performance on \texttt{dbo:WorldHeritageSite} features on a UNESCO-region level, and \texttt{gpt-4-1106-preview} generally performed better on this level as well.

When comparing Table~\ref{tab:eval_results_countries_whsites} and Table~\ref{tab:eval_results_regions_whsites}, we notice greater disparities in the country-level performance than in the UNESCO-region-level performance. The \texttt{gpt-4-1106-preview} model had an accuracy with a range of 0.43 on a country level, compared with a range of 0.133 on a UNESCO-region level. Same for \texttt{gpt-4-vision-preview}, the accuracy had a range of 0.34 on a country level, which was larger than a range of 0.11 on a UNESCO-region level. This means that as the geographic scale increased from countries to UNESCO regions, inter-region disparities in the geo-guessing performance of both models on \texttt{dbo:WorldHeritageSite} features might become smaller.
\begin{table*}[h!]
    \caption{The top ten countries (with more than ten sites) ordered by the percentage of correct predictions by \texttt{gpt-4-1106-preview} and \texttt{gpt-4-vision-preview}, respectively, on \texttt{dbo:WorldHeritageSite} features}
    \centering
    \begin{tabular}{|c|c|} \hline 
           \textbf{gpt-4-1106-preview}& \textbf{gpt-4-vision-preview}\\ \hline 
           France (0.5)& India (0.41)\\ \hline 
           India (0.47)& China (0.33)\\ \hline 
           China (0.39)& Spain (0.31)\\ \hline 
           Italy (0.38)& Italy (0.29)\\ \hline 
           Belgium (0.33)& Belgium (0.25)\\\hline
 Spain (0.31)&Germany (0.23)\\\hline
 Germany (0.23)&France (0.2)\\\hline
 Japan (0.19)&Switzerland (0.17)\\\hline
 Switzerland (0.17)&Japan (0.13)\\\hline
 Chile (0.07)&Chile (0.07)\\\hline
    \end{tabular}
    \label{tab:eval_results_countries_whsites}
\end{table*}

\begin{table*}[h!]
    \caption{The regions (defined by UNESCO for its activities) ordered by the percentage of correct predictions by \texttt{gpt-4-1106-preview} and \texttt{gpt-4-vision-preview}, respectively, on \texttt{dbo:WorldHeritageSite} features}
    \centering
    \begin{tabular}{|c|c|} \hline 
           \textbf{gpt-4-1106-preview}& \textbf{gpt-4-vision-preview}\\ \hline 
           Latin America and the Caribbean (0.413)& Africa (0.37)\\ \hline 
           Asia and the Pacific (0.407)& Asia and the Pacific (0.36)\\ \hline 
           Africa (0.4)& Arab States (0.28)\\ \hline 
           Europe and North America (0.36)& Europe and North America (0.27)\\ \hline 
           Arab States (0.28)& Latin America and the Caribbean (0.26)\\\hline
    \end{tabular}
    \label{tab:eval_results_regions_whsites}
\end{table*}

\section{Conclusions and Future Work}
In this initial work, we explore the notion of geographic diversity through the lens of LLMs, aiming to better understand how well geographic features are represented. In contrast to the common perspective of seeing GPT-4 as a data generator, we also consider it a geographic knowledge base in its own right. We study geographic diversity with a geo-guessing experiment as an open-ended question-answering test, where GPT-4 is utilized to predict a geographic feature masked in a piece of text. Using English-language DBpedia abstracts, we find that GPT-4 may encode insufficient geographic knowledge about several feature types, including \texttt{dbo:WorldHeritageSite}, \texttt{dbo:Valley}, \texttt{dbo:Bay}, and \texttt{dbo:Sea}, on a global level. On a local level, we observe not only this insufficiency but also inter-regional disparities in GPT-4's geo-guessing performance for \texttt{dbo:WorldHeritageSite} features that carry both local and global significance. Interestingly, when assessing on a larger geographic scale, inter-regional disparities may become smaller. Moreover, the multimodal variant of GPT-4 may encode even less geographic knowledge than the unimodal version, whether on a global level for all selected feature types or on a local level for \texttt{dbo:WorldHeritageSite} alone. We speculate that GPT-4 does not perform well in our experiment due to reasons such as the loss in training data compression, the vulnerability to factual contradictions appearing in data conflation, the tendency for LLMs to repeat other named entities (in the prompt) as the correct answer, and so forth. Considering that the training data of GPT-4 is likely to have already included DBpedia, one promising way of enhancing its performance is to implement retrieval-augmented generation~\citep{lewis2020retrieval}, a general-purpose fine-tuning approach that could use DBpedia again as an external knowledge base.

Future work will require a larger-scale but granular analysis of geographic features, supported by various ground-truth knowledge corpora and comprehensive probing techniques. While our experiment provides linguistically and geographically contextual (unstructured) data about a target feature, it is neither a geoparsing task where the feature is unmasked nor a visual GeoGuessr\footnote{\url{https://www.geoguessr.com}} game where a player is asked to locate where a photo was taken. However, these two tasks could give us the inspiration to develop better probing techniques for geographic knowledge. For instance, one could ask LLMs to output a feature name along with geospatial information if representing it with different geometric primitives (e.g., points, lines, polygons), or to list features that are topologically connected if spatial predicates are given. Also, one could replace a masked abstract with their own dataset consisting of multi-perspective descriptions about a geographic feature. In fact, knowledge graphs, such as DBpedia, provide a rich body of structured knowledge, which could help achieve both mainstream probing and conduct our proposed geo-guessing experiment. As knowledge graphs also provide information ontologies, we could study both the \textit{intension} (i.e., the properties of a category) and the \textit{extension} of a geographic feature type and their roles in foundation models.

\section{Acknowledgement}
The authors would like to thank Jiaqi Liang for sharing valuable feedback when forming the initial idea of this work.

\bibliography{reference}

\end{document}